\documentclass[preprint,11pt]{aastex}

\shortauthors{Winn et al.~2009}
\shorttitle{Spin-Orbit Misalignment in XO-3}

\begin{document}

%
\def\ltsima{$\; \buildrel < \over \sim \;$}
\def\lsim{\lower.5ex\hbox{\ltsima}}
\def\gtsima{$\; \buildrel > \over \sim \;$}
\def\gsim{\lower.5ex\hbox{\gtsima}}
                                                                                          
%

\bibliographystyle{apj}

\title{
On the Spin-Orbit Misalignment of the XO-3 Exoplanetary System$^1$
}

\author{
Joshua N.\ Winn\altaffilmark{2},
John Asher Johnson\altaffilmark{3},
Daniel Fabrycky\altaffilmark{4},
Andrew W.\ Howard\altaffilmark{5},\\
Geoffrey W.\ Marcy\altaffilmark{5},
Norio Narita\altaffilmark{6},
Ian J.\ Crossfield\altaffilmark{7},
Yasushi Suto\altaffilmark{8},\\
Edwin L.\ Turner\altaffilmark{9,10},
Gil Esquerdo\altaffilmark{4},
Matthew J.\ Holman\altaffilmark{4}
}

\altaffiltext{1}{Data presented herein were obtained at the W.M.~Keck
  Observatory, which is operated as a scientific partnership among the
  California Institute of Technology, the University of California,
  and the National Aeronautics and Space Administration, and was made
  possible by the generous financial support of the W.~M.~Keck
  Foundation.}

\altaffiltext{2}{Department of Physics, and Kavli Institute for
  Astrophysics and Space Research, Massachusetts Institute of
  Technology, Cambridge, MA 02139, USA}

\altaffiltext{3}{Institute for Astronomy, University of Hawaii,
  Honolulu, HI 96822; NSF Astronomy and Astrophysics Postdoctoral
  Fellow}

\altaffiltext{4}{Harvard-Smithsonian Center for Astrophysics, 60
  Garden Street, Cambridge, MA 02138, USA}

\altaffiltext{5}{Department of Astronomy, University of California,
  Mail Code 3411, Berkeley, CA 94720, USA}

\altaffiltext{6}{National Astronomical Observatory of Japan, Tokyo,
  Japan, Osawa, Mitaka, Tokyo 181-8588, Japan}

\altaffiltext{7}{Department of Astronomy, University of California,
  430 Portola Plaza, Box 951547, Los Angeles, CA}

\altaffiltext{8}{Department of Physics, The University of Tokyo, Tokyo
  113-0033, Japan}

\altaffiltext{9}{Princeton University Observatory, Princeton, NJ,
  08544, USA}

\altaffiltext{10}{Institute for the Physics and Mathematics of the
  Universe, University of Tokyo, Kashiwa 277-8568, Japan}

\begin{abstract}

  We present photometric and spectroscopic observations of the
  2009~Feb.~2 transit of the exoplanet XO-3b. The new data show that
  the planetary orbital axis and stellar rotation axis are misaligned,
  as reported earlier by H\'ebrard and coworkers. We find the angle
  between the sky projections of the two axes to be $37.3\pm 3.7$~deg,
  as compared to the previously reported value of $70\pm 15$~deg. The
  significance of this discrepancy is unclear because there are
  indications of systematic effects. XO-3b is the first exoplanet
  known to have a highly inclined orbit relative to the equatorial
  plane of its parent star, and as such it may fulfill the predictions
  of some scenarios for the migration of massive planets into close-in
  orbits. We revisit the statistical analysis of spin-orbit alignment
  in hot-Jupiter systems. Assuming the stellar obliquities to be drawn
  from a single Rayleigh distribution, we find the mode of the
  distribution to be $13^{+5}_{-2}$~deg. However, it remains the case
  that a model representing two different migration channels---in
  which some planets are drawn from a perfectly-aligned distribution
  and the rest are drawn from an isotropic distribution---is favored
  over a single Rayleigh distribution.

\end{abstract}


\section{Introduction}

Many exoplanets have eccentric orbits. It is widely held that the
orbits were initially circular, due to dissipation in the
protoplanetary disk, and that the eccentricities were somehow excited
after the planets acquired most of their mass. It is also presumed
that orbits were initially aligned with the protoplanetary disk, which
was itself aligned with the equatorial plane of the parent
star. Whether this alignment is generally maintained is not obvious;
whatever mechanism excites the orbital eccentricities may also perturb
the orbital inclinations. For example, large eccentricities may be
produced by close encounters between planets (Rasio \& Ford 1996,
Weidenschilling \& Marzari 1996, Lin \& Ida 1997), which would
occasionally produce large inclinations, even if the initial
inclinations are only a few degrees (Chatterjee et al.~2008).

For close-in giant planets (``hot Jupiters'') in particular, which are
thought to have formed at large orbital distances and then migrated
inward, one may wonder whether the migration process disturbed the
original coplanarity. The various migration theories differ on this
point. Migration via tidal torques from the protoplanetary disk should
not excite the inclination, and may even drive the system toward
closer alignment (Lubow \& Ogilvie~2001). In contrast, migration via
planet-planet scattering would magnify any initial misalignments
(Chatterjee et al.~2008; Nagasawa et al.~2008; Juri\'c \& Tremaine
2008). Scenarios involving Kozai cycles can also leave a highly
inclined final state (Wu et al.~2007; Fabrycky \& Tremaine 2007,
Nagasawa et al.~2008). Thus, one might learn about a planet's
migration history by seeking evidence for a tilt of its orbit with
respect to the stellar equatorial plane.

For transiting planets, spin-orbit alignment is assessed by observing
the Rossiter-McLaughlin (RM) effect, a time-dependent distortion in
the stellar spectral-line profile due to the partial eclipse of the
rotating stellar photosphere. The distortion is usually manifested as
an anomalous Doppler shift during transits. It was first observed in
an exoplanetary system by Queloz et al.~(2000) and has since been
observed in more than a dozen systems. The theory and applications of
the RM effect have also been discussed extensively (Ohta et al.~2005,
Gim\'enez 2006, Gaudi \& Winn 2007, Fabrycky \& Winn 2009).

Analysis of the RM effect allows one to determine $\lambda$, the angle
between the sky projections of the orbital axis and the stellar
rotation axis. All exoplanets that have been examined---with two
exceptions---have been found to be consistent with close alignment and
small values of $\lambda$. One exception was HD~17156b, for which
Narita et al.~(2008) found $\lambda = 62\pm 25$~deg, but follow-up
observations by Cochran et al.~(2008) and Barbieri et al.~(2008)
showed good alignment (as also confirmed by Narita et al., in
prep.). The other exception is XO-3b, for which H\'ebrard et
al.~(2008) (hereafter, H08) found $\lambda=70\pm 15$~deg based on
observations with the 1.93m telescope at the Observatoire de
Haute-Provence (OHP) and the SOPHIE spectrograph (Bouchy et al.~2006).
However, those authors cautioned that additional data were needed to
exclude the possibility that the Doppler measurements were affected by
systematic errors related to the high airmass and bright sky
background of some of their observations.

In this paper we present more definitive data for XO-3, based on
simultaneous spectroscopic and photometric observations of the transit
of 2009~Feb.~2. We describe the observations and data reduction
procedures in \S~2. In \S~3, we present evidence for spin-orbit
misalignment by modeling the RM effect. In \S~4, we discuss the
results and some of their implications.

\section{Observations}

\subsection{Spectroscopy}
\label{subsec:spectroscopy}

We observed the transit of UT~2009~Feb.~2 with the Keck~I 10m
telescope on Mauna Kea, Hawaii. We used the High Resolution Echelle
Spectrometer (HIRES; Vogt et al.~1994) in the standard setup of the
California Planet Search program, as summarized here. We employed the
red cross-disperser and used the I$_2$ absorption cell to calibrate
the instrumental response and the wavelength scale. The slit width was
$0\farcs 86$ and the typical exposure time was 300~s, giving a
resolution of $65,000$ and a signal-to-noise ratio (SNR) of
110~pixel$^{-1}$.

\begin{figure}[ht]
\epsscale{0.9}
\plotone{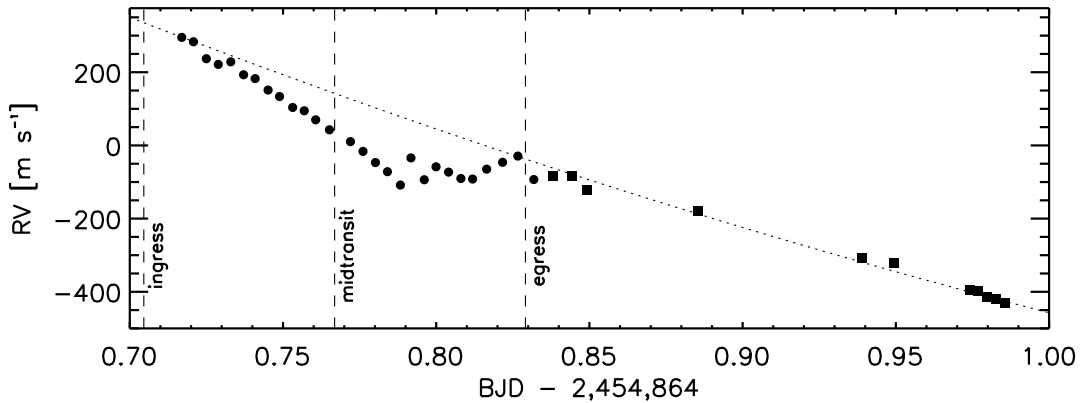}
\caption{ 
{\bf Apparent radial-velocity variation of XO-3 during the 2009~Feb.~2
  transit,} based on observations with Keck/HIRES. The internal
measurement errors are smaller than the symbol sizes. Dashed lines
indicate the photometrically determined times of ingress, midtransit,
and egress. The dotted line is the model of the orbital RV variation
described in \S~\ref{subsec:simple}.
\label{fig:rv}}
\end{figure}

We obtained 39 spectra over 6.5~hr. The observations began just before
$12\arcdeg$ twilight, during the transit ingress. They continued
throughout the 2.5~hr transit and for 0.5~hr after egress. Over the
next 3.5~hr we observed other targets, returning several times to XO-3
to measure the orbital RV variation as precisely as possible. We
determined the relative Doppler shifts with the algorithm of Butler et
al.~(1996). Measurement errors were estimated from the scatter in the
solutions for each 2~\AA~section of the spectrum. The data are given
in Table~\ref{tbl:rv} and plotted in Fig.~\ref{fig:rv}.

\subsection{Photometry}
\label{subsec:photometry}

Simultaneous photometric observations of the 2009~Feb.~2 transit were
conducted with the 1.2m telescope at the Fred L.~Whipple Observatory
(FLWO) in Arizona, and the Nickel 1m telescope at Lick Observatory in
California. At FLWO we used Keplercam, a $4096^2$ CCD with a
$23\arcmin$ square field of view. The images were binned $2\times 2$,
giving a scale of $0\farcs 68$ per binned pixel. We obtained 15~s
exposures though an $r$-band filter for 5.5~hr bracketing the
predicted midtransit time. At Lick Observatory we used the Nickel
Direct Imaging Camera, which has a $2048^2$ CCD with a $6\farcs 3$
square field of view. The images were binned $2\times 2$, giving a
scale of $0\farcs 37$ per binned pixel. We used a Cousins $I$ filter,
and an exposure time between 7--9~s depending on the conditions.

\begin{figure}[p]
\epsscale{0.9}
\plotone{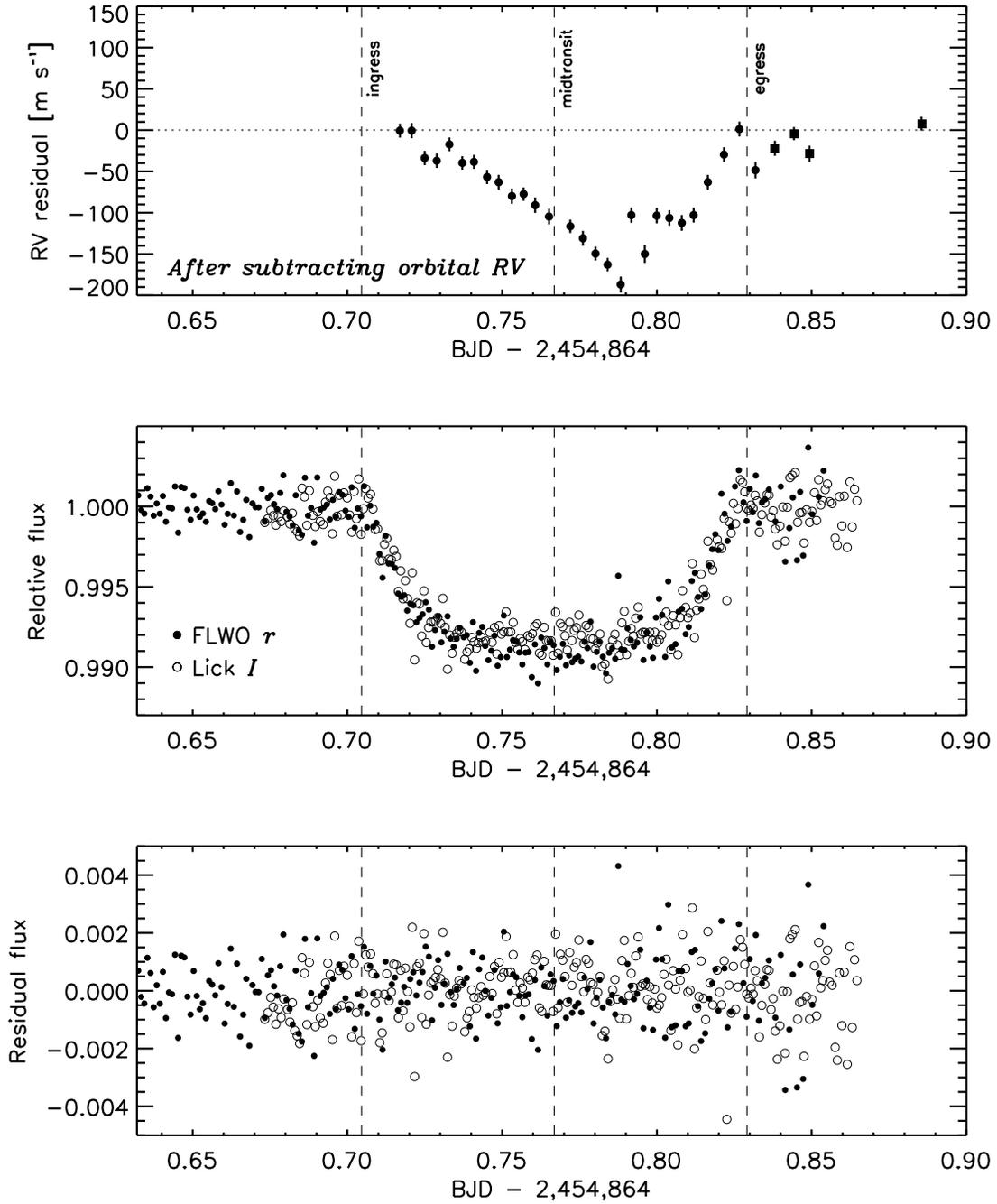}
\caption{ 
{\bf Spectroscopic and photometric observations of XO-3 during the
  2009~Feb.~2 transit.} {\it Top.}---The same data as in Fig.~1, after
subtracting the orbital RV model (see \S~\ref{subsec:simple}). The error
  bars represent internal measurement errors only and do not include
  ``stellar jitter.'' {\it
  Middle.}---Relative photometry based on $r$-band observations with
the FLWO 1.2m telescope and $I$-band observations with the Nickel 1m
telescope. The data have been binned $\times$3 for display purposes.
{\it Bottom.}---Differences between the photometric data
and the best-fitting model.
\label{fig:transit}}
\end{figure}

The CCD images were reduced using standard IRAF\footnote{The Image
  Reduction and Analysis Facility (IRAF) is distributed by the
  National Optical Astronomy Observatory, which is operated by the
  Association of Universities for Research in Astronomy (AURA) under
  cooperative agreement with the National Science Foundation.}
procedures for bias subtraction, flat-field division, and aperture
photometry. The flux of XO-3 was divided by a weighted sum of the
fluxes of comparison stars elsewhere in the field of view. Corrections
were applied to account for systematic effects due to differential
extinction and imperfect flat-fielding, using a procedure described in
\S~3.1. The final time series of relative flux measurements are given
in Table~\ref{tbl:phot}, and plotted in Fig.~\ref{fig:transit}. The
FLWO data have a median time between samples of 29~s and an
out-of-transit standard deviation of 0.0020.  For the Lick data, the
corresponding numbers are 22~s and 0.0022.

\section{Data analysis}

\subsection{Determination of the midtransit time}

As long as a transiting planet's trajectory (the ``transit chord'')
does not pass too close to the center of the stellar disk, a condition
that is fulfilled for XO-3b, the value of $\lambda$ is mainly encoded
in the time difference between the transit midpoint and the moment
when the anomalous Doppler shift vanishes. The main purpose of the new
photometry was to determine the precise midtransit time. The other
photometric parameters, such as the transit duration and depth, have
already been determined by Johns-Krull et al.~(2008) and Winn et
al.~(2008) (hereafter, JK08 and W08) with uncertainties that are
negligible for our purposes.

We fitted a model to each photometric time series in which the free
parameters were the midtransit time $T_c$ and some parameters relating
to systematic effects that were evident in the data. For the FLWO
data, the parameters were $m_0$ and $k_z$ describing a correction due
to differential extinction,
\begin{equation}
m_{\rm cor} = m_{\rm obs} + m_0 - k_zz,
\end{equation}
where $m_{\rm obs}$ is the observed magnitude, $z$ is the airmass, and
$m_{\rm cor}$ is the corrected magnitude that is compared to an
idealized Mandel \& Agol~(2002) model. For the Lick data, a strong
correlation was also found between the out-of-transit flux and the $x$
and $y$ pixel coordinates of XO-3, presumably due to imperfect
flat-field calibration. For this data the correction took the form
\begin{equation}
m_{\rm cor} = m_{\rm obs} + m_0 - k_x x - k_y y - k_z z.
\end{equation}
All of the other relevant parameters were held fixed at the values
determined by W08. To determine the allowed ranges of the parameters,
we used a Markov Chain Monte Carlo (MCMC) algorithm.\footnote{Tegmark
  et al.~(2004), Ford (2005), and Gregory~(2005) provide useful
  background information on this method. For our particular
  implementation see, e.g., Holman et al.~(2006) or Winn et
  al.~(2007a).} For the FLWO data, we assumed the photometric errors to
be Gaussian with a standard deviation equal to the observed
out-of-transit standard deviation. We did the same for the Lick data,
except that we further multiplied the error bars by 1.8 to correct for
time-correlated noise. The factor of 1.8 was determined by examining
the standard deviation of progressively time-binned light curves; for
averaging times between 10 and 20 minutes, the standard deviation
exceeds the expectation of uncorrelated Gaussian noise by a factor of
1.8 (for further discussion see \S~3 of W08).

\begin{figure}[ht]
\epsscale{1.0}
\plotone{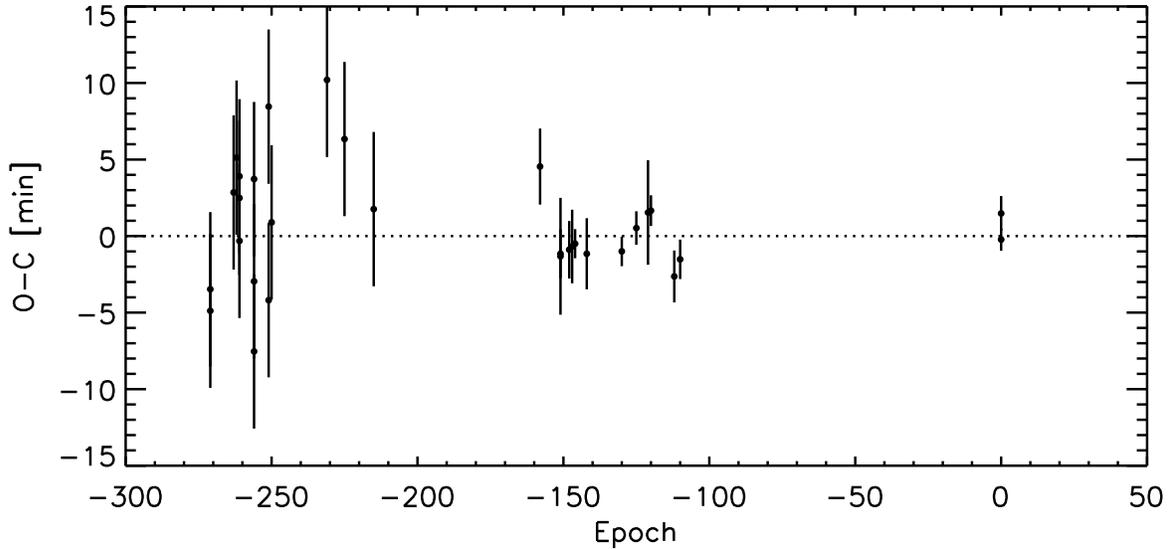}
\caption{
{\bf Transit timing residuals.} A linear function of epoch was fitted
to the transit times of JK08, W08, and this work, and the calculated
times were subtracted from the observed times. The data from the XO
Survey instruments were not used.
\label{fig:ephem}}
\end{figure}

Based on the W08 ephemeris, the predicted midtransit time was
Barycentric Julian Date (BJD) $2454864.7663\pm 0.0010$. The FLWO
result for the midtransit time, expressed in fractional days after
BJD~2454864, is $0.76668\pm 0.00051$. The Lick result is $0.76787\pm
0.00079$. The difference between the FLWO and Lick results is $102\pm
81$~s, suggesting that our error bars are reasonable. We refined the
transit ephemeris by including the two new data points in the
compilation of JK08 and W08 and fitting a linear function $T_c[N] =
T_c[0] + NP$, where $N$ is an integer. The linear fit gave
$\chi^2=30.7$ with 29 degrees of freedom. Fig.~\ref{fig:ephem} shows
the timing residuals. The refined ephemeris is:
\begin{eqnarray}
  T_c[0] & = & 2,454,864.76684 \pm 0.00040~~{\rm BJD} \nonumber \\
  P & = & 3.1915289 \pm 0.0000032~~{\rm days}. \label{eq:ephemeris}
\end{eqnarray}

\subsection{Evidence for spin-orbit misalignment: simple analysis}
\label{subsec:simple}

The apparent radial-velocity variation seen in Fig.~\ref{fig:rv}
arises from both orbital motion and the RM effect. To remove the
variation due to orbital motion and isolate the RM effect, we
determined the parameters of the best-fitting Keplerian orbital model
based on 20 RV measurements published by H08, and then subtracted the
orbital model from the Keck data. Specifically, we used the 19 RVs
gathered at essentially random orbital phases outside of transits
during the few weeks after the transit observation of 28~Jan~2008. We
also used one data point from 28~Jan~2008 that was obtained before the
transit began.

Our MCMC algorithm found the best values and uncertainties of the
velocity semiamplitude $K$, orbital eccentricity $e$, argument of
pericenter $\omega$, systemic velocity $\gamma$, orbital period $P$,
and midtransit time $T_c$. Gaussian prior constraints were imposed on
$P$ and $T_c$ based on Eqn.~(\ref{eq:ephemeris}), and uniform priors
were used for the other parameters. The results were consistent with
the results of H08 but with greater precision in $P$ and $T_c$. We
fixed $\{K,e,\omega,P,T_c\}$ at their optimized values, and found the
choice of $\gamma$ that best fits the out-of-transit Keck data
(indicated by square symbols in Fig.~\ref{fig:rv}). Then we subtracted
the model from the Keck data. The results are shown in the top panel
of Fig.~\ref{fig:transit}.

For a prograde orbit with well-aligned spin and orbital axes, one
expects the RM anomaly to be positive (redshifted) for the first half
of the transit, because the planet is blocking a portion of the
approaching (blueshifted) half of the rotating star. Then one expects
the RM anomaly to vanish at midtransit, when the planet is in front of
the projected stellar rotation axis. Finally, in the last half of the
transit, one expects the RM anomaly to be negative (blueshifted)
because the planet is blocking a portion of the receding (redshifted)
half of the rotating star.

Fig.~\ref{fig:transit} does not show this pattern. Instead, the
anomaly is a blueshift from at least one-quarter of the way into the
transit until its completion. Based on a linear fit to the data from
the first half of the transit, the RM anomaly vanished $\Delta t =
72\pm 9$~min before the transit midpoint. Evidently the planet passed
in front of the projected rotation axis before it reached the midpoint
of the transit chord. This can happen only if the sky-projected
rotation axis and the normal to the transit chord are
misaligned.\footnote{Or equivalently, if the transit chord is
  misaligned with the ``projected stellar equator,'' defined as the
  stellar diameter that is perpendicular to the projected rotation
  axis.} Thus, without detailed modeling of the RM effect we may
conclude that the orbit of XO-3b is inclined with respect to the
rotation axis of its parent star.

If we may approximate the planet's motion across the stellar disk as
uniform and rectilinear, then from the geometry of the transit chord
we may relate $\Delta t$ to $\lambda$:
\begin{equation}
\frac{\Delta t}{T} = \frac{b\tan\lambda}{2\sqrt{1-b^2}},
\end{equation}
where $b$ is the transit impact parameter in units of the stellar
radius and $T$ is the total transit duration, with endpoints defined
by the passage of the center of the planet over the stellar limb.
Using $b$ and $T$ from W08, and $\Delta t = 72\pm 9$~min from the
preceding analysis, we find $\lambda = 44\pm 4$~deg.

Although this calculation has the virtue of simplicity, it is
unsatisfactory in some respects. It relies on an extrapolation to
determine the time when the RM anomaly vanished. The uncertainty in
the orbital model is not taken into account. Also neglected are the
non-uniform motion of the planet across the stellar disk, and the
slight misalignment between the projected orbital axis and the normal
to the transit chord. Furthermore, the information conveyed by
amplitude and duration of the RM waveform is ignored.

\subsection{Evidence for spin-orbit misalignment: comprehensive analysis}
\label{subsec:comprehensive}

A better model includes a simultaneous description of the orbital
motion and the RM effect. The orbital motion is described by a
Keplerian RV curve, as in \S~\ref{subsec:simple}. To model the RM
effect, it is necessary to establish the relation between the
anomalous RV and the configuration of the star and planet. For this
purpose we used the procedure described by Winn et al.~(2005): we
simulated RM spectra with the same format and noise characteristics as
the actual data, and determined the apparent RV using the same
algorithm used on the actual data.

The premise of the simulation is that the only relevant variations in
the emergent spectrum across the visible stellar disk are those due to
uniform rotation and limb darkening. Variations due to differential
rotation, turbulent motion, convective cells, and other effects are
neglected. We begin with a template spectrum with minimal rotational
broadening (described below). We apply a rotational broadening kernel
with $v\sin i_\star=18.5$~km~s$^{-1}$ to mimic the disk-integrated
spectrum of XO-3.\footnote{Here and elsewhere, we use $i_\star$ to
  denote the inclination of the stellar rotation axis with respect to
  the sky plane. This is to be contrasted with $i_o$, the inclination
  of the orbital axis with respect to the sky plane.}  Then we
subtract a scaled, velocity-shifted version of the original
narrow-lined spectrum, intended to represent the portion of the
stellar disk hidden by the planet. We perform this step for many
choices of the scaling $\delta$ and velocity shift $V_p$, and then we
``measure'' the anomalous Doppler shift $\Delta V_R$ of each spectrum.
A polynomial function is fitted to the relation between $\Delta V$ and
$\{\delta, V_p\}$.

The template spectrum should be similar to that of XO-3 but with
comparatively little rotational broadening. We used a Keck/HIRES
spectrum of HD~3861 (F5V, $v\sin i_\star=2.8$~km~s$^{-1}$; Valenti \&
Fischer 2005) with a signal-to-noise ratio of 500 and a resolution of
70,000. Based on the results we adopted the following relation:
\begin{equation}
  \Delta V_R = -\delta~V_p
  \left[
    1.644 - 1.036 \left( \frac{V_p}{{\rm 18.5~km~s}^{-1}}\right)^2
  \right].
\label{eq:deltav}
\end{equation}

The complete RV model is $V_O(t) + \Delta V_R(t)$, where $V_O$ is the
line-of-sight component of the Keplerian orbital velocity and $\Delta
V_R$ is the Rossiter anomaly given by Eq.~\ref{eq:deltav}, with
$\delta$ from the Mandel \& Agol~(2002) model and $V_p$ computed under
the assumption of a uniformly-rotating photosphere. The Keplerian
orbit is parameterized by the period $P$, time of transit $T_c$,
velocity semiamplitude $K$, eccentricity $e$, argument of pericenter
$\omega$, and a constant additive velocity $\gamma$. The RM effect is
parameterized by the projected stellar rotation rate $v\sin i_\star$
and the projected spin-orbit angle $\lambda$.

We fitted the 39 Keck velocities presented in
\S~\ref{subsec:spectroscopy} and the 20 OHP out-of-transit velocities
of H08 that were specified in \S~\ref{subsec:simple}. The OHP
out-of-transit data were included to constrain the Keplerian orbital
parameters. The OHP transit data were {\it not}\, included, thereby
allowing the Keck data to provide an independent determination of the
transit parameters $\lambda$ and $v\sin i_\star$. The OHP and Keck
data were granted independent values of $\gamma$.

We determined the credible intervals for the model parameters with an
MCMC algorithm. Uniform priors were used for $K$, $e\cos\omega$,
$e\sin\omega$, $\gamma_{\rm OHP}$, $\gamma_{\rm Keck}$, $v\sin
i_\star$, and $\lambda$. Gaussian priors were used for $P$ and $T_c$,
based on the ephemeris of Eq.~(\ref{eq:ephemeris}). For the parameters
needed to compute the fractional loss of light $\delta$, namely the
transit depth, total duration, and duration of ingress or egress, we
used Gaussian priors based on the results of W08.\footnote{An
  additional parameter with a minor role is the linear limb-darkening
  coefficient $u$ that is used to compute $\delta$. We set $u=0.5$
  based on the expectation for a star such as XO-3 in the red optical
  band (Claret~2004). Varying this parameter by as much as $\pm 0.3$
  or even allowing it to be a free parameter makes no essential
  difference in the results.} We found it necessary to enlarge the RV
errors by adding 14.1~m~s$^{-1}$ in quadrature with the measurement
errors in order to achieve a reduced $\chi^2$ of unity. This is a
plausible level of intrinsic velocity noise (``stellar jitter'') for
an F5V star, based on the empirical findings of Wright~(2005).

\begin{figure}[p*]
\epsscale{0.8}
\plotone{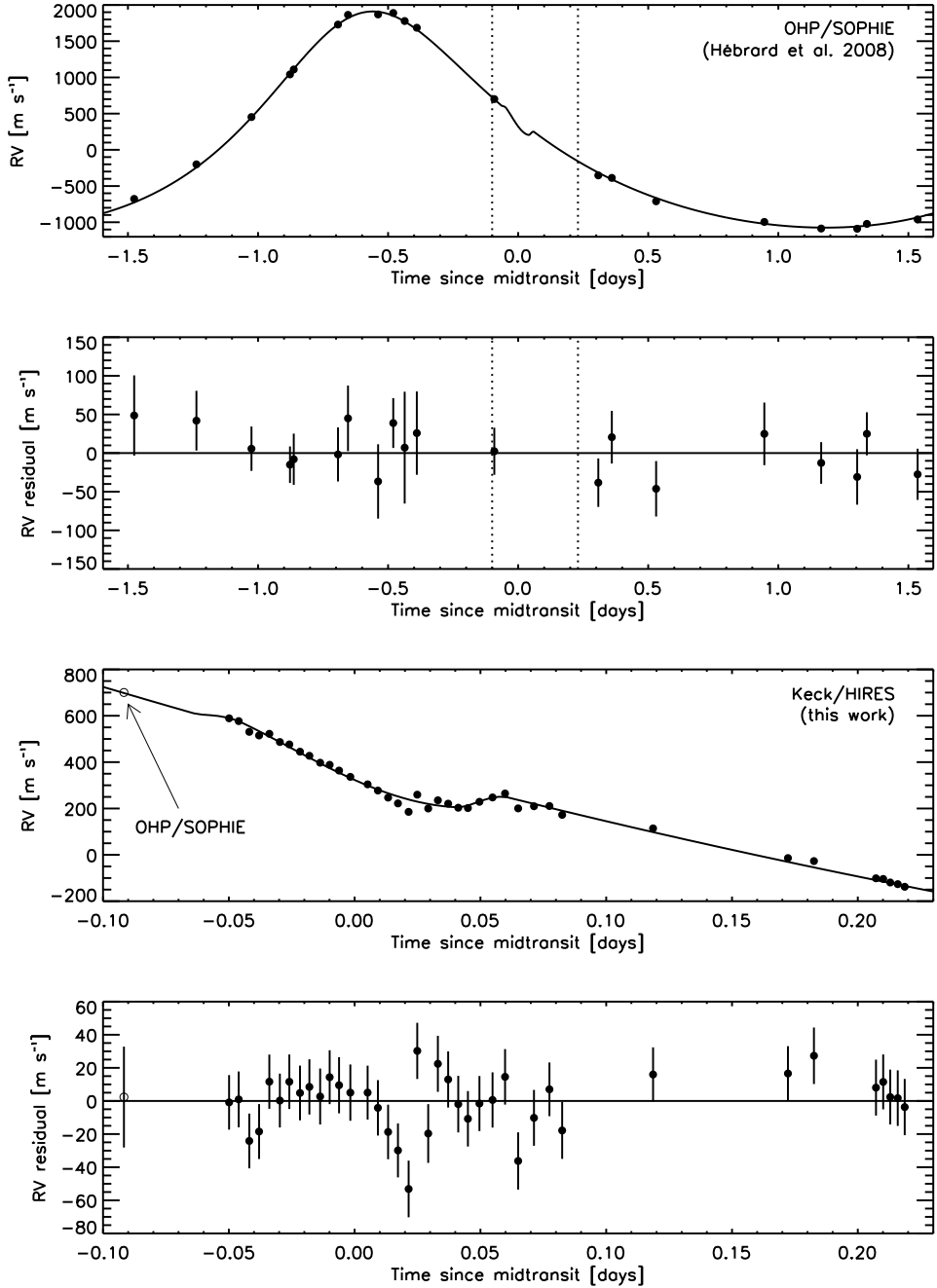}
\caption{
{\bf Radial-velocity model including orbital motion and the RM
  effect.} (a) The H08 data. The solid curve is the best-fitting
model. The dotted lines indicate the time range plotted in the lower 2
panels. (b) Differences between the H08 data and the best-fitting
model. (c) The Keck/HIRES data from the transit of 2009~Feb.~2. Also
plotted is the single point from H08 that was used in the
model-fitting procedure. (d) Differences between the Keck/HIRES data
and the best-fitting model. The error bars are the quadrature
sums of the internal measurement error and 14.1~m~s$^{-1}$ (the ``stellar jitter'' term).
\label{fig:model}}
\end{figure}

The results for the model parameters are given in
Table~\ref{tbl:params}. The quoted value for each parameter is the
median of the {\it a posteriori}\, distribution, marginalized over all
other parameters. The quoted 1$\sigma$ (68.3\% confidence) errors are
defined by the 15.85\% and 84.15\% levels of the cumulative
distribution. Our results for $K$, $e$, $\omega$, and $\gamma_{\rm
  OHP}$ are in accord with the previous analysis of H08, the only
differences having arisen from our choice of priors on $P$ and $T_c$.
As expected, the results for the photometric parameters were very
close to the Gaussian prior constraints that were imposed. The
best-fitting model is shown in Fig.~\ref{fig:model}.

Through this analysis we found $\lambda = 37.3\pm 3.7$~deg. A similar
result was obtained in \S~3.2 using a na\"{\i}ve model, although we
have already noted some shortcomings of the simple analysis. The
comprehensive model takes into account the nonuniform orbital
velocity, as well as all other relevant information and uncertainties
in the external parameters. It also gives an independent estimate of
the projected stellar rotation rate, $v\sin i_\star = 18.31\pm
1.3$~km~s$^{-1}$. There is good agreement with the value determined
from the stellar line-broadening, $18.54\pm 0.17$~km~s$^{-1}$ (JK08),
which is a consistency check on our interpretation of the RV data and
our model of the RM effect.

\section{Discussion}

Through simultaneous photometry and spectroscopy of a transit of
XO-3b, we have determined that the planetary orbit is significantly
inclined relative to the equatorial plane of its host star. According
to our analysis, the angle between the sky projections of the orbital
axis and the stellar rotation axis is $\lambda = 37.3\pm 3.7$~deg.
Thus we have confirmed the finding of H08 that XO-3b is the first
known case of an exoplanet whose orbit is highly inclined with respect
to the equatorial plane of its parent star, or equivalently, the first
exoplanetary system in which the host star is known to have a large
obliquity.

Quantitatively, our result for $\lambda$ disagrees with the previous
result ($70\pm 15$~deg) by 2.1$\sigma$, where $\sigma$ is the
quadrature sum of the errors in the two results. The significance of
this discrepancy is small and its interpretation is unclear. It is
possible the systematic effects over which H08 expressed concern, due
to high airmass and moonlight, have biased the result for $\lambda$.
It is also possible we are underestimating the error in $\lambda$ by
assuming the RV errors to be Gaussian and uncorrelated. The Keck RV
residuals do not appear to be Gaussian and uncorrelated; in particular
the 4 largest outlying data points were all from the narrow time range
between 0.02--0.04~d after midtransit. The noise may include artifacts
of the instrument or data reduction procedures, or sources of apparent
radial-velocity variation besides Keplerian orbital motion and the RM
effect---such as star spots, other activity-induced variations, and
additional planets---that are correlated on the time scale of the
transit. The best way to reduce this source of uncertainty is to
gather more spectroscopic transit data, preferably covering an entire
transit and including plenty of pre-ingress and post-egress data.

Because the angle $i_\star$ is unknown, the projected spin-orbit angle
$\lambda= 37.3\pm 3.7$~deg gives a lower limit on the true angle
$\psi$ between the orbital axis and the stellar rotation axis. Thus
the orbit of XO-3b is more inclined relative to its host star than any
planet in the Solar system, including Pluto ($\psi=12.2$~deg).

The unknown angle $i_\star$ is also relevant to the determination of
the stellar rotation period $P_{\rm rot}$. Using the estimates of
$R_\star$ and $v\sin i_\star$ by W08 and JK08,
\begin{equation}
P_{\rm rot} = \frac{2\pi R_\star}{v\sin i_\star}~\sin i_\star = (3.73\pm 0.23~{\rm d})~\sin i_\star.
\end{equation}
This is not far from the orbital period of 3.19~d and thus it is
possible that the spin and orbit are synchronized, or
pseudo-synchronized (H08). However, tidal theory would predict that
such a system would have a fleeting existence, for the following
reasons. Tidal evolution does not lead to a stable equilibrium
configuration for most of the star-planet pairs, possibly including
XO-3 (Rasio et al.~1996, Levrard et al.~2009). Even if the star were
tidally spun up, it would be expected to lose angular momentum through
a wind and consequently consume the planet (Barker \& Ogilvie
2009). Furthermore, the dissipation in the star that would drive its
spin into pseudo-synchonization with the orbit would also damp the
orbital eccentricity on a similar timescale, because the ratio of
orbital to spin angular momentum is small (Hut 1981). Hence, the
observation of a significant eccentricity suggests that the stellar
spin has not been significantly altered by tides. However, some other
systems present circumstantial evidence for tidal spin-orbit
interactions (McCullough et al.~2008, Pont~2008).

The large obliquity of its host star is not the only unusual property
of XO-3b. Even by the standards of hot Jupiters it is an outlier, with
an unusually large mass ($12$~M$_{\rm Jup}$) and orbital eccentricity
($e=0.29$). Naturally one wonders if there is a connection between
these properties and the nonzero value of $\lambda$, as mentioned in
the introduction. There are at least two other massive planets on
close-in eccentric orbits which have been subjects of RM observations,
and in both cases $\lambda$ has been found to be consistent with zero:
HAT-P-2b (Winn et al.~2007b, Loeillet et al.~2008) and HD~17156b
(Cochran et al.~2008, Barbieri et al.~2008).\footnote{Moutou et
  al.~(2009) recently reported observations of the RM effect in the
  highly eccentric HD~80606 system. The authors did not determine
  $\lambda$ because their data only cover the latter part of the
  transit, but they found that the data are compatible with a large
  spin-orbit misalignment if the orbital inclination is not too close
  to $90\arcdeg$.} Is XO-3b anomalous even within the subgroup of
close-in massive planets on eccentric orbits?  The answer is not yet
clear. It must be remembered that the RM effect is sensitive only to
the projected spin-orbit angle and it is therefore possible that the
other systems are also significantly misaligned; this is especially so
for HAT-P-2b because the small impact parameter of the transit
degrades the achievable precision in $\lambda$.

Fabrycky \& Winn~(2009) presented a framework for overcoming the
problem of projection effects using a Bayesian analysis of the results
from many different planets. They found that the conclusions that
could be drawn from the current ensemble were strongly driven by the
case of XO-3b. Specifically, they compared two descriptions of the
data: (1) a model in which a fraction $f$ of systems have perfect
spin-orbit alignment and the rest have random mutual orientations; (2)
a model in which the spin-orbit angle $\psi$ is drawn from a Rayleigh
distribution (or more precisely a Fisher distribution on a sphere,
which reverts to a Rayleigh distribution when it is highly
directional). They calculated the Bayesian evidence
\begin{equation}
E \equiv \int p({\rm data} | \alpha) p(\alpha) d\alpha
\end{equation}
for each model, where $\alpha$ is the single free parameter of the
model (either the fraction $f$, or the mode $\sigma$ of the Rayleigh
distribution). They found $E$ to be 134 times greater for the first
model, a finding that was interpreted as evidence for two distinct
modes of planet migration. However, they cautioned that this
conclusion hinged on the tentative finding of $\lambda=70\pm 15$~deg
for XO-3.

We have repeated this analysis using the revised estimate of $\lambda=
37.3\pm 3.7$~deg and making no other changes. Results from the first
model, a division between perfectly-aligned and randomly-oriented
systems, remain nearly the same: $f<0.36$ with 95\% confidence, and
the Bayesian evidence $E$ is 1920 (having fallen only slightly from
1927). Results from the second model changed more significantly. The
reduced value of $\lambda$ for XO-3 makes a single Rayleigh
distribution more probable, with $E=69$ instead of 14. In addition,
with the improved precision of the new result, $\lambda=0$ is ruled
out with higher confidence, leading to a sharper lower limit on the
mode $\sigma$ of the Rayleigh distribution. We find
$\sigma=13_{-2}^{+5}$~deg (1$\sigma$ errors). In this model, the most
probable value of $\psi$ among hot Jupiter systems is larger than the
value of 6~deg between the Solar rotation axis and Jupiter's orbital
axis. The first model is preferred, with a formal confidence of
$E_1/(E_1+E_2) = 96.6\%$. However, the confidence has been reduced
from the value of 99.28\% calculated by Fabrycky \& Winn (2009). Thus,
although the evidence for two wholly distinct modes of planet
migration has been weakened, it is still highly suggestive.

As with many attempts to derive conclusions based on {\it a
  posteriori}\, statistics, a problem with the foregoing analysis is
that subtle and ill-quantified selection effects have shaped the
sample under consideration. In particular, XO-3b was not selected
randomly for this study. The previous finding of a large value of
$\lambda$ was a motivating factor that has led to improved precision
in $\lambda$, and consequently greater weight in the statistical
analysis. Furthermore, the idea to fit a model consisting of two
distributions (perfectly-aligned and isotropic) was developed only
after knowing of XO-3's possibly strong misalignment.

Apart from those thorny issues, the high sensitivity of the
statistical results to a single data point means that the results must
be treated with caution, and underlines the importance of gathering
additional data. Observing more transits of XO-3b is advisable, given
the possibility noted earlier of correlated RV noise. Observations of
the Rossiter-McLaughlin effect are also desired for other transiting
systems, spanning a range of planetary masses, orbital eccentricities,
and orbital periods. Such observations would elucidate any connections
between the planetary and orbital parameters, and may provide
important clues about the processes that lead to close-orbiting
planets.

\acknowledgments We thank Peter McCullough for reminding us about the
near-coincidence of the orbital and rotational periods of XO-3b. We
are grateful to G\'asp\'ar Bakos for trading telescope time at FLWO on
short notice, and to Breann Sitarski and Tracy Ly for helping with the
observations at Lick Observatory. We are indebted to Lara Winn for
enabling this work to be completed in a timely fashion. This work was
partly supported by the NASA Origins program through awards NNX09AD36G
and NNX09AB33G. JAJ acknowledges support from an NSF Astronomy and
Astrophysics Postdoctoral Fellowship (grant no.\ AST-0702821).  This
work was partly supported by World Premier International Research
Center Initiative (WPI Initiative), MEXT, Japan.


\begin{deluxetable}{lcc}

\tabletypesize{\scriptsize}
\tablecaption{Relative Radial Velocity Measurements of XO-3\label{tbl:rv}}
\tablewidth{0pt}

\tablehead{
\colhead{BJD} &
\colhead{RV [m~s$^{-1}$]} &
\colhead{Error [m~s$^{-1}$]}
}

\startdata
  $  2454864.71696$  &  $    295.28$  &  $   8.47$  \\
  $  2454864.72077$  &  $    283.24$  &  $   9.22$  \\
  $  2454864.72498$  &  $    236.89$  &  $   8.63$  \\
  $  2454864.72887$  &  $    221.36$  &  $   8.68$  \\
  $  2454864.73296$  &  $    228.46$  &  $   8.29$  \\
  $  2454864.73714$  &  $    193.07$  &  $   8.10$  \\
  $  2454864.74090$  &  $    182.65$  &  $   8.45$  \\
  $  2454864.74513$  &  $    151.37$  &  $   8.58$  \\
  $  2454864.74887$  &  $    133.62$  &  $   8.85$  \\
  $  2454864.75315$  &  $    103.70$  &  $   9.28$  \\
  $  2454864.75693$  &  $     94.67$  &  $   8.10$  \\
  $  2454864.76066$  &  $     69.94$  &  $   9.39$  \\
  $  2454864.76514$  &  $     42.68$  &  $   9.49$  \\
  $  2454864.77202$  &  $     10.35$  &  $   8.11$  \\
  $  2454864.77610$  &  $    -16.32$  &  $   8.98$  \\
  $  2454864.78016$  &  $    -46.80$  &  $   8.49$  \\
  $  2454864.78408$  &  $    -71.76$  &  $   8.12$  \\
  $  2454864.78831$  &  $   -108.26$  &  $   9.64$  \\
  $  2454864.79175$  &  $    -34.11$  &  $   9.40$  \\
  $  2454864.79610$  &  $    -93.81$  &  $  10.74$  \\
  $  2454864.79994$  &  $    -58.44$  &  $   9.33$  \\
  $  2454864.80403$  &  $    -72.96$  &  $   9.37$  \\
  $  2454864.80804$  &  $    -90.44$  &  $   9.62$  \\
  $  2454864.81188$  &  $    -91.84$  &  $   9.02$  \\
  $  2454864.81646$  &  $    -64.90$  &  $   8.79$  \\
  $  2454864.82167$  &  $    -46.12$  &  $   8.89$  \\
  $  2454864.82664$  &  $    -29.13$  &  $   9.05$  \\
  $  2454864.83185$  &  $    -93.29$  &  $   9.95$  \\
  $  2454864.83813$  &  $    -84.01$  &  $   9.20$  \\
  $  2454864.84428$  &  $    -83.04$  &  $   7.95$  \\
  $  2454864.84934$  &  $   -121.17$  &  $   9.76$  \\
  $  2454864.88547$  &  $   -179.66$  &  $   8.36$  \\
  $  2454864.93909$  &  $   -307.71$  &  $   8.57$  \\
  $  2454864.94942$  &  $   -320.68$  &  $   9.57$  \\
  $  2454864.97407$  &  $   -394.97$  &  $   9.24$  \\
  $  2454864.97694$  &  $   -397.80$  &  $   8.71$  \\
  $  2454864.97973$  &  $   -413.02$  &  $   8.65$  \\
  $  2454864.98273$  &  $   -420.18$  &  $   9.08$  \\
  $  2454864.98550$  &  $   -431.56$  &  $   9.46$  

\enddata

\tablecomments{The RV was measured relative to an arbitrary template
  spectrum; only the differences are significant. The uncertainty
  given in Column 3 is the internal error only and does not account
  for any possible ``stellar jitter.''}

\end{deluxetable}


\begin{deluxetable}{cccc}

\tabletypesize{\scriptsize}
\tablecaption{Relative Photometry of XO-3\label{tbl:phot}}
\tablewidth{0pt}

\tablehead{
\colhead{Observatory\tablenotemark{a}} &
\colhead{BJD} &
\colhead{Rel.~Flux} &
\colhead{Error}
}

\startdata
  $  1$  &  $  2454864.62513$  &  $  0.9995$  &  $  0.0020$ \\
  $  1$  &  $  2454864.62544$  &  $  1.0011$  &  $  0.0020$ \\
  $  1$  &  $  2454864.62579$  &  $  0.9997$  &  $  0.0020$ \\
  $  1$  &  $  2454864.62612$  &  $  1.0008$  &  $  0.0020$ \\
  $  1$  &  $  2454864.62646$  &  $  1.0015$  &  $  0.0020$ \\
  $  1$  &  $  2454864.62678$  &  $  0.9985$  &  $  0.0020$ \\
  $  1$  &  $  2454864.62712$  &  $  1.0007$  &  $  0.0020$ \\
  $  1$  &  $  2454864.62743$  &  $  1.0006$  &  $  0.0020$
\enddata

\tablenotetext{a}{(1) FLWO 1.2m telescope, $r$ band. (2) Nickel 1m
  telescope, $I$ band.  We intend for this Table to appear in entirety
  in the electronic version of the journal. An excerpt is shown here
  to illustrate its format.}

\end{deluxetable}


\begin{deluxetable}{lcc}

\tabletypesize{\normalsize}
\tablecaption{Radial-Velocity Model Results for XO-3\label{tbl:params}}
\tablewidth{0pt}

\tablehead{
\colhead{Parameter} &
\colhead{Value} &
\colhead{Uncertainty}
}

\startdata
Projected spin-orbit angle, $\lambda$~[deg]    &  $37.3$  & $3.7$ \\
Projected stellar rotation rate, $v \sin i_\star$~[km~s$^{-1}$]   &  $18.31$ &  $1.3$ \\
Velocity semi-amplitude, $K$~[m~s$^{-1}$]  &  $1488$  & $10$ \\
Orbital eccentricity, $e$             &  $0.2884$ &  $0.0035$ \\
Argument of pericenter, $\omega$~[deg]  &  $346.3$ &  $1.3$ \\
Velocity offset, $\gamma_{\rm Keck}$~[m~s$^{-1}$]         &  $-293.6$ &  $7.0$ \\
Velocity offset, $\gamma_{\rm OHP}$~[m~s$^{-1}$]         &  $-12045.4$ &  $8.0$
\enddata

\end{deluxetable}


\end{document}